\documentclass[10pt, conference, letterpaper]{IEEEtran}
\IEEEoverridecommandlockouts
\usepackage[scaled=0.92]{helvet}
\usepackage{times}
\usepackage{multirow}
\usepackage{graphicx,subfigure}
\usepackage{algorithm}
\usepackage{algorithmic}
\usepackage{amsthm}
\usepackage{amsmath}
\usepackage{color}
 \theoremstyle{definition}    \newtheorem{lm}{Theorem}
\theoremstyle{plain}         \newtheorem{theorem}[lm]{Theorem}
\theoremstyle{definition}    
\theoremstyle{plain}         
\theoremstyle{definition}    
\theoremstyle{plain}         
\graphicspath{{./fig/}}
\usepackage{cite}
\usepackage{url}
\usepackage{textcomp}
\usepackage{amssymb}

\newcounter{thm} 

\DeclareMathOperator{\argmax}{argmax}

\newcounter{exampleI}
\newcounter{exampleII}
\newcounter{exampleIII}
\newcommand {\commentout}[1] {}

\def\bull{\vrule height .9ex width .8ex depth -.1ex }

\newcommand {\beq} {\begin{equation}}
\newcommand {\eeq} {\end{equation}}
\newcommand {\bear} {\begin{eqnarray}}
\newcommand {\eear} {\end{eqnarray}}
\newcommand {\barr} {\begin{array}}
\newcommand {\earr} {\end{array}}
\newcommand {\Ns} {{\rm I\kern-2.5pt N}}
\newcommand {\Z} {{\rm I\kern-2.5pt Z}}
\newcommand {\Rs} {{\rm I\kern-2.5pt R}}

\newenvironment{Proofwithheader}[1]{\medbreak
\noindent
{\bf Proof #1:~}}{\unskip\nobreak\hfill\hskip 2em \bull \par\medbreak}

\newcommand{\eat}[1]{}

\begin{document}

\title{\huge Mutual or Unrequited Love: Identifying Stable Clusters in
  Social Networks with  Uni- and Bi-directional Links}
\author{Yanhua Li, Zhi-Li Zhang\thanks{The work is supported in part by the NSF grants CNS-0905037,
CNS-1017647 and the DTRA Grant HDTRA1-09-1-0050.} and Jie Bao
\\
Department of Computer Science \& Engineering, University of Minnesota, Twin Cities\\
Email: \{yanhua,zhzhang,baojie\}@cs.umn.edu}


\maketitle
\begin{abstract}

Many social networks, e.g., Slashdot and Twitter, can be
represented as directed graphs ({\em digraphs}) with two types of links between entities:
mutual (bi-directional) and one-way (uni-directional) connections. Social science theories reveal that mutual
connections are more stable than one-way connections, and
one-way connections exhibit various tendencies to become mutual
connections. It is therefore important to take such tendencies into
account when performing clustering of social networks with both mutual
and one-way connections.

In this paper, we utilize the {\em dyadic} methods to analyze social
networks, and develop a generalized mutuality tendency theory to
capture the tendencies of those node pairs which tend to establish
mutual connections more frequently than those occur by chance. Using
these results, we develop a {\em mutuality-tendency-aware} spectral
clustering algorithm to identify more stable clusters by maximizing
the {\em within-cluster} mutuality tendency and minimizing the {\em
cross-cluster} mutuality tendency. Extensive simulation results on
synthetic datasets as well as real online social network datasets
such as Slashdot, demonstrate that our proposed
mutuality-tendency-aware spectral clustering algorithm extracts more
stable social community structures than traditional spectral
clustering methods.

\end{abstract}

\vspace{2mm}
\section{Introduction}\label{sec:intro}
\vspace{2mm}

 Graph models are widely utilized to represent relations
among entities in social networks. Especially, many online social
networks, e.g., Slashdot and Twitter, where the users' social
relationships are represented as directed edges in directed graphs
(or in short, \emph{digraphs}). Entity connections in a digraph can
be categorized into two types, namely, bi-directional links (mutual
connections) and uni-directional links ({\em one-way} connections).
Social theories \cite{wolfe1997social} and online social network
analysis~\cite{wolfe1997social, descioli2011best,jamali2011modeling}
have revealed that various types of connections exhibit different
stabilities, where mutual connections are more stable than one-way
connections. In other words, mutual connections are the source of
social cohesion~\cite{gouldner1960norm,golder2009structural} that,
if two individuals mutually attend to one another, then the bond is
reinforced in each direction.

Studying the social network structure and properties of social ties
have been an active area of research. Clustering and identifying
social structures in social networks is an especially important
problem
~\cite{kurucz2007spectral,smyth2005spectral,mishra2007clustering}
that has wide applications, for instance, community detection and
friend recommendation in social networks. Existing clustering
methods~\cite{icmlZhouHS05,edbt:SatuluriP11} are originally
developed for {\em undirected} graphs, based on the classical {\em
spectral clustering theory}. Several recent studies (see, e.g.,
~\cite{edbt:SatuluriP11,icmlZhouHS05,
WestonLIZEN05,leicht2008community}) extend the spectral clustering
method to digraphs, by first converting the underlying digraphs to
undirected graphs via some form of {\em symmetrization}, and then
apply spectral clustering to the resulting symmetrized (undirected)
graphs. However, all these methods
have two common drawbacks, which prevent them from obtaining {\em
stable} clusters with {\em more mutual connections}. First, these
methods do not explicitly distinguish between  {\em mutual} and {\em
one-way} connections commonly occurring in many social networks,
treating them essentially as the same and therefore ignoring the
different social relations and interpretations  these two types of
connections represent (see Section~\ref{sec:prelim} for more
in-depth discussion).
\begin{figure}[htb]
\centering \vspace{-0mm}
\includegraphics[width=0.5 \textwidth]{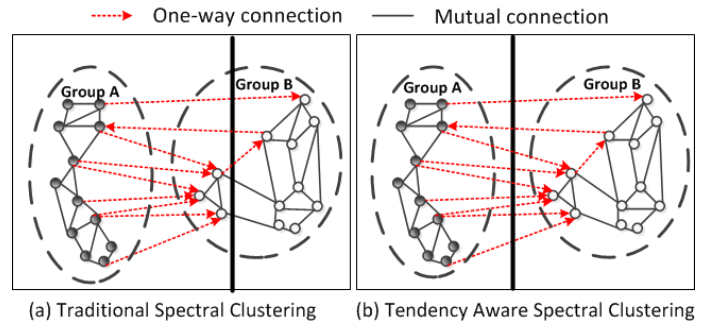} \vspace{-3mm}
  \caption{An example network}\label{intro}
\vspace*{-0.0cm}
\end{figure}
Second, by simply minimizing the total cross-cluster links (that are
symmetrized in some fashion),  these methods do not explicitly account
for the potential tendencies of node pairs to become mutually connected.
As a simple example,  Fig.~\ref{intro} shows two groups of people in a
network, where
people in the same group tend to have more mutual (stable)
connections, and people across two groups have more one-way
(unstable) connections. When using the traditional spectral
clustering method, as shown in Fig.~\ref{intro}(a), group B will be
partitioned into two clusters, due to its strict rule of minimizing
the total number of across cluster edges. On the other hand, the
correct partition should be done as shown in Fig.~\ref{intro}(b),
where the majority of mutual (stable) connections are placed within
clusters, and one-way (unstable) connections are placed
across clusters.

In this paper, we propose and develop a stable social cluster detection
algorithm that takes into account the tendencies of node
pairs whether to form mutual (thus stable) connections or not, which
can result in more \emph{stable} cluster structures. To tackle this clustering
problem, we need to answer the following questions: 1) how to track
and evaluate the tendencies of node pairs to become mutual (stable)
relations? and 2) how to cluster the entities in social networks by
accounting for their mutuality tendencies so as to extract more stable
clustering structures?

To address these questions, we utilize  dyadic methods to analyze social
networks, and develop a generalized mutuality tendency theory which better
captures the tendencies of node pairs that tend to establish
mutual connections  more frequently than those occur by chance. Using
these results, we
develop a {\em mutuality-tendency-aware} spectral clustering algorithm
to detect more stable clusters by maximizing the
{\em within-cluster} mutuality
tendency and minimizing the {\em cross-cluster}
mutuality tendency. Our contributions are
summarized as follows.
\begin{itemize}

  \item Motivated by the social science mutuality tendency theory,
  we establish a new {\em cluster-based} mutuality tendency theory
  which yields a symmetrized mutuality tendency for each node
  pair, and provides a measure of strength of social ties among nodes in a
  cluster.

  \item Based on our theory, we develop a
    \emph{mutuality-tendency-aware} spectral
  clustering algorithm that can partition the social graphs into stable
  clusters, by maximizing the within-cluster mutuality tendencies
  and minimizing the across cluster mutuality tendencies.

  \item The experimental results -- based on both social network
  structures of synthetical and real social network datasets -- confirm that our
  clustering algorithm is able to generate more stable clusters
  than the traditional spectral clustering algorithms.

\end{itemize}

 To our best knowledge, this is the first work studying the
  impact of tendencies of node pairs to become mutual connections
  on the stability of cluster structure of social networks. The
  remainder of the paper is organized as follows. In
  section~\ref{sec:prelim} we  briefly discuss the existing dyadic
  analysis methods, the traditional spectral clustering
  algorithms and other related work. In section~\ref{sec:mutuality} we
  introduce a cluster-base mutuality tendency theory, and based on
  this theory, we develop a mutuality-tendency-aware spectral clustering
  algorithm in section~\ref{sec:ourmethod}. In section~\ref{sec:eva},
  we evaluate the performances of our method using synthetic  and real
  social network (e.g., Slashdot) datasets. We conclude the paper in
  section~\ref{sec:conclusion}.

\vspace{2mm}
\section{Preliminaries, Related Work and Problem definition}\label{sec:prelim}
\vspace{0.2cm}

In this section, we first introduce the existing dyadic analysis
methods in the social theory literature for analyzing and
characterizing social network mutual connections and one-way
connections. We then present the classic spectral clustering theory
which was developed for {\em undirected} graphs, and briefly survey
some related works which apply this theory to {\em digraphs} through
{\em symmetrization}. We argue that these existing methods for
clustering digraphs via symmetrization are inadequate in solving
social network clustering problems, as they ignore different social
ties (and mutuality tendencies) represented by mutual and one-way
connections in social networks. We end the section with the problem
definition, namely, how to identify {\em stable} clusters in social
networks by taking into account mutuality tendencies of mutual and
one-way connections.


%

\subsection{Dyadic Analysis and Mutuality Tendency}
Given a social network with both uni- and bi-directional links, such
a network can be represented as a (simple) digraph $G=(V,E)$ with
$|V|=n$ nodes. If the links also have weights (say, representing the
strength of connections or social ties), such a network can be more
generally represented as a {\em weighted} digraph, $G=(V,E,A)$ where
$A_{ij}$ represents the strength of connection or  ``affinity'' from
node $i$ to node $j$. When $A$ is a 0-1 matrix, $G$ reduces to a
simple digraph, and $A$ is the standard adjacency matrix of the
digraph, {\color{black}where $A_{ij}=1$ if the directed edge
$i\rightarrow j$ is present, and $A_{ij}=0$ otherwise}. In this
paper for simplicity we focus primarily on simple (unweighted)
digraphs {\color{black} with no selfloops}, namely, social networks
with unweighted directional links. Most online social networks are
of unweighted variety.

Social scientists commonly view the social network $G$ as
a collection of dyads~\cite{wolfe1997social}, where \emph{a dyad is
an {\em unordered} pair of nodes and {\em directed} edges between two nodes in
the pair.} Denote a dyad as $Dy_{ij}=(A_{ij}, A_{ji})$, for $i < j$.
Since dyad is an unordered notion, we have in total $N_d=n(n-1)/2$ dyads in $G$.
Hence, there are only three possible isomorphism dyads. The first type of dyads
is {\em mutual} relationship, where both directional edges $i\rightarrow
j$ and $j\rightarrow i$ are present. The second type of dyads is
{\em one-way} relationship, where either $i\rightarrow j$
or $j\rightarrow i$ is present, but not both. The last type of dyads
is {\em null} relationship, where no edges show up between $i$ and $j$.

\noindent\textbf{Interpretations of dyads.}
Social scientists have observed that mutual
social relations and one-way relations in social networks typically
exhibit different stabilities, namely, mutual relations are more stable
than one-way relations~\cite{wolfe1997social}. Hence in the social
science literature, one prevalent interpretation of  dyadic relations in social networks are the following:
mutual dyads are considered as stable connections between two nodes
and null relation dyads represent no relations; the one-way
dyads~\cite{heider1946attitudes,berscheid2005psychology,miller1972structural,price1966psychological,rodrigues1967effects}
are viewed as an {\em intermediate} state of relations, which are in
transition to more stable equilibrium states of reciprocity (mutual or
no relation).  Several recent empirical
studies~\cite{kwak2011fragile,jamali2011modeling} of online social
networks have further revealed and confirmed  that  mutual social
relations are more stable relations than one-way connections.


%

\noindent\textbf{Computing dyad census.} Given a (simple) digraph
$G=(V,E)$, with $n=|V|$ nodes. Let $m$, $b$, and $u$ denote the
number of mutual, one-way, and null dyads in the network. Clearly,
$m+b+u=n(n-1)/2$. The triple $\langle m,b,u \rangle$ is referred to
as the {\em dyad census}, since it is derived from an examination of
all (possible) dyads in the network.
The dyad census triple can be computed in terms of the adjacency
matrix $A$ of $G$ as follows (in both scalar and matrix forms):
\begin{align}
&m=\sum_{i<j}A_{ij}A_{ji}=\frac{1}{2}\textbf{tr}(AA),\nonumber \\
&b=|E| -2m= \textbf{tr}(AA^T)-\textbf{tr}(AA), \nonumber \\
&u=N_d-b-m=N_d-\textbf{tr}(AA^T)+\textbf{tr}(AA). \nonumber
\end{align}

\noindent\textbf{Measuring mutuality tendency.} The notion of
mutuality tendency has been introduced in the social science
literature (see, e.g., ~\cite{katz1955measurement,wolfe1997social})
to measure the tendency for a node pair to establish mutual
connections.  For any dyad between $i$ and $j$ in a  digraph $G$, if
$i$ places a link to $j$, $\rho_{ij}$ represents the tendency that
$j$ will reciprocate to $i$ more frequently than would occur by
chance.

{\color{black} Let $\textbf{X}_{ij}$ denote the random variable that
represents whether or not node $i$ places a directed edge to node
$j$. There are only two possible events (i.e., $\textbf{X}_{ij}$
takes two possible values): $\textbf{X}_{ij}=1$, representing the
edge is present; or $\textbf{X}_{ij}=0$, the edge is not present.
Let $X_{ij}$ (resp. $\bar{X}_{ij}$) denote the event
$\{\textbf{X}_{ij}=1\}$ (resp. $\{\textbf{X}_{ij}=0\}$). Then the
probability of the event $X_{ij}$ occurring is $P(X_{ij})$. The
probability that $i$ places a directed edge to $j$ and $j$
reciprocates back (i.e., node $i$ and node $j$ are mutually
connected) is thus given by
\begin{align}
&P(X_{ij},X_{ji})=P(X_{ij})P(X_{ji}|X_{ij}),
\label{eq:allprobability}
\end{align}
Wofle~\cite{wolfe1997social} introduces the following measure
of mutuality tendency in terms of the conditional probability
$P(X_{ji}|X_{ij})$ as  follows:
\begin{align}
&P(X_{ji}|X_{ij})=P(X_{ji})+ \rho_{ij} P(\bar{X}_{ji}), \nonumber\\
&\rho_{ij} =
\frac{P(X_{ij},X_{ji})-P(X_{ij})P(X_{ji})}{P(X_{ij})P(\bar{X}_{ji})},
\label{eq:exp}
\end{align}}
where $-\infty < \rho_{ij}\leq 1$ ensures $0\leq P(X_{ji})+ \rho
P(\bar{X}_{ji})\leq 1$ to hold. Like many indices used in
statistics, $\rho_{ij}$ is dimensionless and easy to interpret,
since it uses $0$ and $1$ as benchmarks. If $\rho=1$, the mutuality
tendency is maximum, meaning that given that node $i$ places a link
to node $j$, node $j$ will for sure reciprocate. If $\rho_{ij}=0$
(i.e., $P(X_{ji}|X_{ij})= P(X_{ji})$), then node $j$ reciprocates
and places a link to node $i$ purely by chance, namely, it is
independent of the event that node $i$ places a link to node $j$.
Hence when  $0< \rho_{ij} \leq 1$, it suggests more than a chance
tendency for node $j$ to reciprocate back. Furthermore, if
$\rho_{ij}<0$ (i.e., $P(X_{ji}|X_{ij})< P(X_{ji})$), there is less
than chance tendency for node $j$ to reciprocate; in other words, it
suggests a tendency away from mutual dyads, toward one-way and null
dyads.  Hence, $-\infty < \rho_{ij}\leq 1$ provides a measure of the
strength of tendency for reciprocation.

{\color{black}From eq.(\ref{eq:exp}), the joint distribution
$P(X_{ij},X_{ji})$ in eq.(\ref{eq:exp}) can be measured by the
observed graph, namely, either
$P(X_{ij},X_{ji})=P^{(\omega)}(X_{ij},X_{ji})=1$, when $i$ and $j$
have mutual connection, or
$P(X_{ij},X_{ji})=P^{(\omega)}(X_{ij},X_{ji})=0$, otherwise, where
the superscript $\omega$ indicates that the probability is obtained
from the observed graph. On the other hand, the distribution for
each individual edge is measured by
$P(X_{ij})=P^{(\mu)}(X_{ij})=\frac{d_i}{|V|-1}$, where $d_i$ is the
out-going degree of node $i$. $P^{(\mu)}(X_{ij})$ represents the
probability of edge $i\rightarrow j$ being generated under a random
graph model, denoted by the superscript $\mu$, with edges randomly
generated while preserving the out-degrees. Hence, the tendency
$\rho_{ij}$ is obtained by implicitly comparing the observed graph
with a reference random digraph model.}

\noindent\textbf{Limitations of Wolfe's mutuality tendency measure
for stable social structure clustering.}
Although the node pair in a dyad is unordered (i.e., the two nodes are
treated ``symmetrically'' in terms of dyadic relations), Wolfe's
measure of mutual tendency is in fact {\em asymmetric}. This can be
easily seen through the following derivation. By definition,
\begin{align}
&P(X_{ji}|X_{ij})=P(X_{ji})+ \rho_{ji} P(\bar{X}_{ji}), \nonumber \\
&P(X_{ij}|X_{ji})=P(X_{ij})+ \rho_{ij} P(\bar{X}_{ij}). \nonumber
\end{align}
Multiplying the above two equations with $P(X_{ij})$ and $P(X_{ji})$
respectively and from eq.(\ref{eq:allprobability}),
 we have
\begin{align}
&\frac{\rho_{ji}}{\rho_{ij}}  = \frac{P(X_{ji})
P(\bar{X}_{ij})}{P(X_{ij}) P(\bar{X}_{ji})} =
\frac{P(X_{ji})-P(X_{ij})P(X_{ji})}{P(X_{ij})-P(X_{ij})P(X_{ji})}
\nonumber
\end{align}
We see that $\rho_{ij} =\rho_{ji}$ if and only if
$P(X_{ij})=P(X_{ji})$ holds. 
Hence, given an arbitrary dyad in a social network Wolfe's measure
of mutuality tendency of the node pair is asymmetric -- in a sense
that it is a {\em node-specific} measure of mutuality tendency. It
does not provide a measure of mutuality tendency of the (unordered)
{\em node pair} viewed together. While such asymmetric ({\em
  node-specific}) measure of mutuality tendency can be useful in some
social network analysis, as will be clear later, such an asymmetric
measure poses difficulty in identifying and extracting {\em stable}
cluster structures in social networks. For instance, given a
partition $V=(S,\bar{S})$ of a digraph, generalizing Wolfe's measure
to clusters, the mutuality tendencies  {\em across} the two
clusters, denoted by $\rho(S,\bar{S})=\sum_{i\in S, j\in
\bar{S}}\rho_{ij}$ and $\rho(\bar{S}, S)=\sum_{i\in \bar{S},
j\in{S}}\rho_{ij}$, are generally not symmetric, namely,
$\rho(S,\bar{S}) \neq \rho(\bar{S}, S)$. In
Section~\ref{sec:mutuality}, we will introduce a new measure of
mutuality tendency that is {\em symmetric} and captures the tendency
of a node pair in a dyadic relation to establish mutual connection.
This measure of mutuality tendency can be applied to clusters and a
whole network in a straightforward fashion, and leads us to develop
a \emph{mutuality-tendency-aware} spectral clustering algorithm.


\subsection{Spectral Clustering Theory and Extensions to Digraphs via Symmetrization}

Spectral clustering methods (see, e.g., \cite{
Luxburg:2007,pami:ShiM00,WangD:KDD10,icmlZhouHS05,WestonLIZEN05}) are originally developed  for clustering data with
symmetric relations, namely, data that can be represented as {\em
  undirected} graphs, where each relation (edge) between two entities,
$A_{ij}=A_{ji}$, represents their similarity. The goal is to
partition the graph such that entities within each cluster are more
similar to each other than those across clusters. This is done by
minimizing the total weight of cross-cluster edges (possibly
weighted by the total weight of edges within clusters). In the
following we present the basics of spectral clustering theory (see
~\cite{Tutorial-SC}  for more details).

Given the (non-negative) similarity  matrix $A$, the cut
function is defined to quantatively measure the quality of a
partition $V=(S_1, \cdots, {S}_K)$, and is defined as follows:
\begin{align}
Cut(S_l, \bar{S}_l) &:= \sum_{i\in S_l,j\in \bar{S}_l} A_{ij},
\nonumber \\
Cut(S_1, \cdots, {S}_K) &:= \sum_{i=1}^K Cut(S_i,\bar{S}_i).
\nonumber
\end{align}
To account for cluster sizes -- especially to obtain relatively
balanced clusters (in terms of sizes),  the ratio cut function
$RCut$~\cite{hagen1992new} and the normalized cut function
$NCut$~\cite{pami:ShiM00} have also been defined:
\begin{align}
RCut(S_1,\cdots,S_K) &:= \sum_{i=1}^K \frac{Cut(S_i,\bar{S}_i)}{|S_i|}, \nonumber \\
NCut(S_1,\cdots,S_K) &:= \sum_{i=1}^K \frac{Cut(S_i, \bar{S}_i)}{Vol(S_i)}, \nonumber
\end{align}
where $vol(S_i)=\sum_{j\in S_i} d_{j}$ is the volume of the cluster $S_i$.

In the following (and the remainder of the paper),
 we will use the ratio cut function as the objective
function. All the results also hold true for the normalized cut.
Using the ratio cut, the clustering problem formulated as
      a graph mincut optimization problem
 can be rewritten in the following form:
\begin{align}
\min_{S_1,\cdots,S_k} RCut(S_1,\cdots,S_K), \label{eq:min}
\end{align}

The (unnormalized) Laplacian matrix $L=D-A$ is used to solve the
above mincut problem, where $D=\emph{diag}[d_i]$ with
$d_i=\sum_j{A_{ij}}$ is the diagonal degree matrix. Given a
(nonnegative) symmetric $A$, $L$ is symmetric and positive semi-definite. If we take $K$ eigenvectors corresponding
to the smallest eigenvalues of $L$, the optimal solution
to the problem eq.(\ref{eq:min}), namely, the
optimal partition into $K$ clusters, can be well approximated by
applying the K-means algorithm to clustering
 the data points projected to the
subspace formed by these $K$ eigenvectors~\cite{Tutorial-SC}.
Moreover,~\cite{leskovec2010empirical} provides a systematic study
on comparing a wide range of undirected graph based clustering
algorithms using real large datasets, which gives a nice guideline
of how to select clustering algorithms based on the underlying
networks and the targeting objectives.

\noindent\textbf{Extensions to digraphs via symmetrization.} When
relations between entities are {\em asymmetric}, or the underlying
graph is {\em directed}, spectral clustering cannot be directly
applied, as the notion of (semi-)definiteness is only defined for
{\em
  symmetric} matrices. Several recent studies (see, e.g.,
~\cite{edbt:SatuluriP11,icmlZhouHS05,
WestonLIZEN05,leicht2008community}) all attempt to circumvent this
difficulty by first converting the underlying digraphs to undirected
graphs via some form of {\em symmetrization}, and then apply
spectral clustering to the resulting symmetrized (undirected)
graphs. For example, the authors in ~\cite{edbt:SatuluriP11} discuss
several symmetrization methods, including the symmetrized adjacency
matrix $\bar{A}=(A+A^T)/{2}$, the bibliographic coupling matrix
$AA^T$ and the co-citation strength matrix $A^TA$, and so forth.
Symmetrization can also be done through a random walk on the
underlying graph, where $P =D^{-1}A$ is the probability transition
matrix and $D =diag [d^{out}_i]$ is a diagonal matrix of node
out-degrees. For example, taking the objective function as the
random walk flow circulation matrix $F_{\pi}=\Pi P$, where $\Pi$ is
the diagonal stationary distribution matrix, we have the symmetrized
Laplacian of the circulation matrix as
\begin{align}
\bar{\cal L}&=\frac{\tilde{\cal L}+\tilde{\cal
L}^T}{2}=I-\frac{\Pi^{\frac{1}{2}}{P}\Pi^{-\frac{1}{2}}+\Pi^{-\frac{1}{2}}{P^T}\Pi^{\frac{1}{2}}}{2}.\nonumber
\end{align}
where $\tilde{\cal L}$ is the (asymmetric) digraph Laplacian
matrix~\cite{WAW2010Yanhua}. Then the classical spectral clustering
algorithm can then  be applied using $\bar{\cal L}$ which is
symmetric and semi-definite. Zhou and et al~\cite{icmlZhouHS05,
WestonLIZEN05} use this type of symmetrization to perform clustering
on digraphs. {\color{black}Moreover, Leicht and
Newman~\cite{leicht2008community} propose the digraph modularity
matrix $Q=[Q_{ij}]$, which captures the difference between the
observed digraph and the hypothetical random graph with edges
randomly generated by preserving the in- and out-degrees of nodes,
namely, $Q_{ij}=A_{ij}-d_i^{out}d_j^{in}/m$. Then, if the sum of
edge modularities in a cluster $S$ is large, nodes in $S$ are well
connected, since the edges in $S$ tend to appear with higher
probabilities than occur by chance. However, $Q$ by definition is
asymmetric, where~\cite{leicht2008community} uses the symmetrized
$\bar{Q}=(Q+Q^T)/2$ as objective to perform spectral clustering
method. Essentially, the edge modularity captures how an individual
edge appears more frequently than that happens by chance, thus the
modularity based clustering method tends to group those nodes with
more connections than expected together, which like all other
clustering methods presented above completely ignores the
distinction between mutual and one-way connections.} 

\noindent\textbf{Problem definition: Clustering and identifying
stable clusters in
  social networks with mutual and one-way connections.}
As discussed earlier, one-way and mutual dyadic connections in
social networks often represent different states or types of social
ties and exhibit various stabilities over time. Hence when
performing clustering to extract community structures in social
networks, one-way and mutual connections should be distinguished and
treated differently. Existing digraph clustering methods via
symmetrization, e.g., those mentioned above, on the other hand,
ignore these different types of connections and treat them as the
same: the process of symmetrization essentially weighs one-way
connections as a fraction of  mutual connections, and then attempt
to minimize the total weight of the (symmetrized) cross-cluster
links. {\color{black}Moreover, different from Leicht and Newman's
~\cite{leicht2008community} reference random graph model, as
presented in earlier section, the mutuality tendency compares the
observed the digraph with a random graph model where edges are
randomly generated by preserving only the out-degrees, which better
reflects the underlying model of how social
network users establish social ties.} 

In this paper we want to solve the following clustering problem in
social networks with bi- and uni-directional links: Given a directed
(social) graph where mutual connections represent more stable
relations and one-way connections represent intermediate
transferring states, {\em how can we account for mutual tendencies
of dyadic relations and  cluster the entities in such a way that
nodes within each cluster have maximized mutuality tendencies to
establish mutual connections, while across clusters, nodes  have
minimized tendencies to establish mutual connections?} The clusters
(representing social structures or communities) identified and
extracted thereof will hence likely be more stable.



\vspace{-0mm}
\section{Cluster-based Mutuality Tendency
Theory}\label{sec:mutuality} \vspace{-0mm}


Inspired by Wolfe's study in~\cite{wolfe1997social}, we propose a
new measure of mutuality tendency for dyads that can be generalized
to groups of nodes (clusters), and develop a {\em mutuality tendency
theory} for characterizing the strength of social ties within a
cluster (network structure) as well as across clusters in an
asymmetric social graph. This theory lays the theoretical foundation
for the network structure classification and community detection
algorithms we will develop in section~\ref{sec:ourmethod}.

\subsection{Cluster based mutuality tendency}

Let $\textbf{X}_{ij}$ denote the random variable that represents
whether or not node $i$ places a directed edge to node $j$. 
There are only two possible events (i.e., $\textbf{X}_{ij}$ takes
two possible values): $\textbf{X}_{ij}=1$, representing the edge is
present; or $\textbf{X}_{ij}=0$, the edge is not present. Let
$X_{ij}$ (resp. $\bar{X}_{ij}$) denote the event
$\{\textbf{X}_{ij}=1\}$ (resp. $\{\textbf{X}_{ij}=0\}$). Given an
{\em observed} (asymmetric) social graph $G$, to capture the {\em
mutuality tendency} of dyads in this graph, we compare it with a
{\em hypothetical}, {\em random} (social) graph, denoted as
$G^{(\mu)}$, where links (dyadic relations) are generated randomly
(i.e., by chance) in such a manner that the (out-)degree $d_i$ of
each node $i$ in $G^{(\mu)}$ is the same as that in the observed
social graph $G$. Under this random social graph model, the
probability of the event $X_{ij}$ occurring is
$P^{(\mu)}(X_{ij})=\frac{d_i}{|V|-1}$; namely, $i$ places a
(directed) link to node $j$ randomly or by chance (the superscript
$\mu$ indicates the probability distribution of link generations
under the random social graph model).  The probability that $i$
places a directed edge to $j$ and $j$ reciprocates back (i.e., node
$i$ and node $j$ are mutually connected) is thus given by
$P^{(\mu)}(X_{ij},X_{ji})=P^{(\mu)}(X_{ij})P^{(\mu)}(X_{ji}|X_{ij})=P^{(\mu)}(X_{ij})P^{(\mu)}(X_{ji})$,
since $\textbf{X}_{ij}$ and $\textbf{X}_{ji}$ are independent under
the random social graph model. On the observed social graph, denote
$P^{(\omega)}(X_{ij},X_{ji})$ to represent the event whether there
is a mutual connection (symmetric link) between node $i$ and node
$j$, i.e., $P^{(\omega)}(X_{ij},X_{ji})=1$, if the dyad $Dy_{ij}$ is
a mutual dyad in the {\em observed} social graph, and
$P^{(\omega)}(X_{ij},X_{ji})=0$, otherwise.
We define the {\em
  mutuality tendency} of dyad $Dy_{ij}$ as follows:
\begin{align}
\small
\theta_{ij}:&=P^{(\omega)}(X_{ij},X_{ji})-P^{(\mu)}(X_{ij},X_{ji})\nonumber
\\
&=P^{(\omega)}(X_{ij},X_{ji})-P^{(\mu)}(X_{ij})P^{(\mu)}(X_{ji}),\label{eq:tendency}
\end{align}
which captures how the node pair $i$ and $j$ establish a mutual dyad
more frequently than would occur by chance.

This definition of mutuality tendency is a symmetric measure for
dyad $Dy_{ij}$,
  i.e., $\theta_{ij}=\theta_{ji}$. In addition, it is shown that
  $\theta_{ij}\in [-1,1]$. We remark that
  $\theta_{ij}=0$ indicates that if node $i$ places a directed
  link to node $j$, the tendency that node $j$ will reciprocate back to
  node $i$ is no more likely than would occur by chance; the same
  holds true if node $j$ places a directed link to node $i$
  instead. On the other hand, $\theta_{ij}> 0$ indicates that if node
  $i$ (resp. node $j$) places a directed
  link to node $j$ (resp. node $i$), node $j$ (resp. node $i$) will
  more likely than by chance to  reciprocate. In particular, with
  $\theta_{ij}= 1$, node $j$ (resp. node $i$) will almost surely
  reciprocate.   In contrast,  $\theta_{ij}< 0$ indicates that if node
  $i$ (resp. node $j$) places a directed
  link to node $j$ (resp. node $i$), node $j$ (resp. node $i$) will
  tend not to  reciprocate back to node $i$ (resp. node $j$). In
  particular, with $\theta_{ij}= -1$, node $j$ (resp. node $i$) will
  almost surely not reciprocate back.
  Hence $\theta_{ij}$ provides a measure of strength of social ties
  between node $i$ and $j$: $\theta_{ij} >0$ suggests that the dyadic
  relation between node $i$ and $j$ is stronger, having a higher
  tendency (than by chance) to become mutual; whereas $\theta_{ij} <0$
  suggests that node $i$ and $j$ have weaker social ties, and their
  dyadic relation is likely to remain asymmetric or eventually disappear.


\noindent{\textbf{Mutuality tendency of clusters.}} 
The mutuality tendency measure for dyads defined in
eq.(\ref{eq:tendency}) can be easily generalized for an arbitrary
cluster (a subgraph) in an observed social graph, $S\subseteq G$. We
define the mutuality tendency of a cluster $S$, $\Theta_S$, as
follows:

\begin{align}
\small &\Theta_{S}:=\sum_{i\sim j; i,j\in S}P^{(\omega)}(X_{ij},
X_{ji})-\sum_{i\sim j; i,j\in S}P^{(\mu)}(X_{ij}, X_{ji})\nonumber
\\
&=\sum_{i\sim j; i,j\in S}P^{(\omega)}(X_{ij}, X_{ji})-\sum_{i\sim
j; i,j\in S}P^{(\mu)}(X_{ij})P^{(\mu)}(X_{ji}),\label{eq:tendency-S}
\end{align}
\noindent{}where the subscript $i\sim j: i,j\in S$ means that the
summation accounts for all (unordered) dyads, and $i, j$ are both in
$S$. Denote the second term in eq.(\ref{eq:tendency-S}) as
$m^{(\mu)}_S$, and the (out-degree) volume of the cluster $S$ as
$d_S:= \sum_{i\in S} d_i$. As $P^{(\mu)}(X_{ij})=d_i/(|V|-1)$ and
$P^{(\mu)}(X_{ji})=d_j/(|V|-1)$,
\begin{align}
\small m^{(\mu)}_S &=\sum_{i\sim j; i,j\in
S}\frac{d_id_j}{(|V|-1)^2} = \frac{d^2_S -\sum_{i\in
      S}d^2_i}{2(|V|-1)^2} \label{eq:random-m_S},
\end{align}
which represents the expected number of mutual connections among
nodes in $S$ {\em under the random social
  graph model}. Given the cluster $S$ in the observed social graph
$G$, define $m^{(\omega)}_S:= \sum_{i\sim j; i,j\in
  S}P^{(\omega)}(X_{ij}, X_{ji})$, namely, $m^{(\omega)}_S$ represents
the number of (observed) mutual connections among nodes in the
cluster $S$ in the observed social graph $G$. The mutual tendency of
cluster $S$ defined in eq.(\ref{eq:tendency-S}) is therefore exactly
$\Theta_{S}= m^{(\omega)}_S-m^{(\mu)}_S$.

Hence $\Theta_{S}$ provides a measure of strength of (likely mutual)
social ties among nodes in a cluster: $\Theta_{S}>0$ suggests that
there are more  mutual connections among nodes in $S$ than would
occur by chance; whereas $\Theta_{S}<0$ suggests that there are
fewer mutual connections among nodes in $S$ than would occur by
chance. Using $\Theta_{S}$, we can therefore quantify and detect
clusters of nodes (network structures or communities) that have
strong social ties.

In particular, when $S=G$, $\Theta_G$ characterizes the mutuality
tendency for the entire digraph $G$, i.e.,
\begin{align}
\small &\Theta_{G}= m^{(\omega)}_G-m^{(\mu)}_G=\sum_{i\sim
j}\theta_{ij},\label{eq:tendency-G}
\end{align}
where $m^{(\omega)}_G:= \sum_{i\sim j}P^{(\omega)}(X_{ij}, X_{ji})$
represents the number of (observed) mutual dyads among nodes in the
observed social graph $G$, and
\begin{align}
\small m^{(\mu)}_G &=\sum_{i\sim j}\frac{d_id_j}{(|V|-1)^2} =
\frac{d^2 -\sum_{i\in V}d^2_i}{2(|V|-1)^2}, \label{eq:random-m_G}
\end{align}
represents the expected number of mutual dyads among nodes in $G$
{\em under the random social graph model}. Likewise, given a
bipartition $(S, \bar{S})$ of $G$, we define the cross-cluster
mutuality tendency as
\begin{align}
\Theta_{\partial S}&:=
\sum_{i\in S\sim j\in \bar{S}}(P^{(\omega)}(X_{ij}
X_{ji})-P^{(\mu)}(X_{ij})P^{(\mu)}(X_{ji}))\label{eq:tendency-partialS}
\end{align}
Denote the second quantity in eq.(\ref{eq:tendency-partialS}) as
$m^{(\mu)}_S$,
\begin{align} \small m^{(\mu)}_{\partial S} &=\sum_{i\in S\sim j\in
\bar{S}}\frac{d_id_j}{(|V|-1)^2}= \frac{d_Sd_{\bar{S}}}{(|V|-1)^2}
\label{eq:random-m_partialS}
\end{align}
which represents the expected number of mutual connections among
nodes across $S$ and $\bar{S}$ under the random social graph model.
Define $m^{(\omega)}_{\partial S}:= \sum_{i\in S\sim j\in \bar{S}}
P^{(\omega)}(X_{ij}, X_{ji})$ representing the number of (observed)
mutual connections among nodes across clusters $S$ and $\bar{S}$ in
the observed social graph $G$. The mutuality tendency across cluster
$S$ and $\bar{S}$ defined in eq.(\ref{eq:tendency-partialS}) is
therefore exactly $\Theta_{\partial S} = m^{(\omega)}_{\partial S}
-m^{(\mu)}_{\partial S}$.


The mutuality tendency theory outlined above accounts for different
interpretations and roles mutual and one-way connections represent
and play in asymmetric social graphs, with the emphasis in
particular on the importance of mutual connections in forming and
developing stable social structures/communities with strong social
ties. In the next section, we will show how we can apply this
mutuality tendency theory for detecting and clustering stable
network structures and communities in asymmetric social graphs.

\vspace{-0mm}
\section{Mutuality-tendency-aware spectral clustering
algorithm}\label{sec:ourmethod} \vspace{-0mm} In this section, we
first consider the simpler case of mutuality-tendency-aware
clustering problem with $K=2$ and establish the basic theory and
algorithm. We then extend it to the general case with $K> 2$.
\subsection{Mutuality-tendency-aware spectral clustering: K=2}


Without loss of generality, we consider only simple (unweighted)
digraphs $G=(V,E)$ (i.e., the adjacency matrix $A$ is a 0-1 matrix).
Define the mutual connection matrix  $M:=\min(A,A^T)$, which
expresses all the mutual connections with unit weight $1$. In other
words, if node $i$ and node $j$ are mutually connected (with
bidirectional links), $M_{ij}=M_{ji}=1$, otherwise,
$M_{ij}=M_{ji}=0$. Hence, we have $M_{ij}=P^{(\omega)}(X_{ij},
X_{ji})$, representing the event whether there is a mutual
connection (symmetric link) between node $i$ and node $j$, i.e., in
the dyad $Dy_{ij}$ in the observed social graph. In addition, let
$\delta_{ij}$ be the Kronecker delta symbol, i.e., $\delta_{ij}=1$
if $i=j$, and $\delta_{ij}=0$ otherwise. Then, we define matrix
\begin{align}
\bar{M}&=\frac{dd^T - \emph{diag}[d^2]}{(|V|-1)^2} \nonumber
\end{align}
with $d$ as the out-going degree vector, where each entry
\begin{align}
\bar{M}_{ij}&=\frac{d_id_j-\delta_{ij}d_i^2}{(|V|-1)^2}=\left\{
\begin{array}{ll}
    \frac{d_id_j}{(|V|-1)^2} &\textrm{if $i\neq j$} \\
    0 &\textrm{if $i=j$}
\end{array}
\right.\label{eq:Mb}
\end{align}
represents the probability that two nodes $i$ and $j$ independently
place two unidirectional links to each other to form a mutual dyad.
Hence, $\bar{M}_{ij}=P^{(\mu)}(X_{ij})P^{(\mu)}(X_{ji})$ represents
the probability of node pair $i$ and $j$ to establish a mutual
connection under random graph model with edges randomly generated by
preserving the node out-degrees. We denote $T=M-\bar{M}$ as the
mutuality tendency matrix, with each entry
\begin{align}
T_{ij}&=P^{(\omega)}(X_{ij},X_{ji})-P^{(\mu)}(X_{ij})P^{(\mu)}(X_{ji})=\theta_{ij}
\label{eq:TendMatrix}
\end{align}
as the individual dyad mutuality tendency.

\noindent{\textbf{Mutuality Tendency Lapacian}.} $T$ is symmetric
and those entries associated with non-mutual dyads are negative,
representing less mutuality tendencies to establish mutual
connections than those occur by chance. Define the mutuality
tendency Laplacian matrix as
\begin{align}
& L_T = D_T - T \label{eq:TendLap}
\end{align}
where $D_T=diag[d_T(i)]$ is the diagonal degree matrix of $T$, with
$d_T(i)=\sum_{j}T_{ij}$. We have the following theorem presenting
several properties of $L_T$.
\begin{theorem}[]\label{thm:Laplacian}
The mutuality tendency Laplacian matrix $L_T$ as defined in
eq.(\ref{eq:TendLap}) has the following properties
\begin{itemize}
  \item Given a column vector $x \in \mathbb{R}^{|V|}$, the bilinear form $x^T L_Tx$
  satisfies
\begin{align}
& x^T L_T x = \sum_{i\sim j}T_{ij}(x_i-x_j)^2.\label{eq:bil}
\end{align}
  \item $L_T$ is symmetric and in general indefinite. In addition, $L_T$ has one eigenvalue equal to $0$, with corresponding
eigenvector as $\textbf{1}=[1,\cdots,1]^T$.
\end{itemize}
\end{theorem}
\begin{Proofwithheader}{}
(1) By expanding the bilinear form $x^T L_T x$,
\begin{align}
 x^T L_T x &= \sum_{i,j\in
V}T_{ij}(x_i^2-x_ix_j)=
\sum_{i\sim j}T_{ij}(x_i-x_j)^2\nonumber
\end{align}

(2) The symmetry of both $M$ and $\bar{M}$ in
eq.(\ref{eq:TendMatrix}) insures the symmetry of $L_T$, thus $L_T$
has all real eigenvalues. However, $L_T$ is in general indefinite,
because $T_{ij}$ in eq.(\ref{eq:bil}) could be either positive or
negative. On the other hand, since $\textbf{1}^TL_T=\textbf{0}^T$
and $L_T\textbf{1}^T=\textbf{0}$ hold true, $L_T$ has an eigenvalue
equal to $0$ with corresponding eigenvectors as identity vector
$\textbf{1}=[1,\cdots,1]^T$.
\end{Proofwithheader}

\noindent\textbf{Mutuality tendency ratio cut function.}
{\color{black}For a digraph $G=(V,E)$, and a partition
$V=(S,\bar{S})$ on $G$, we define the {\em mutuality tendency ratio
cut function} as follows.
\begin{align}
& TRCut(S, \bar{S}) = \Theta_{\partial S}
\left(\frac{1}{|S|}+\frac{1}{|\bar{S}|}\right), \label{eq:TendCutF}
\end{align}
which represents the overall mutuality tendency across clusters
balanced by the ``sizes'' of the clusters.} 
Then, the clustering problem is formulated as a minimization problem
with $K=2$ clusters. (More general cases with $|V|\geq K>2$ will be
discussed in the next subsection.)
\begin{align}
&\min_{S} TRCut(S, \bar{S})\label{eq:SP-P}
\end{align}
Since $\Theta_{\partial S} = \Theta_{G} - (\Theta_{S} +
\Theta_{\bar{S}})$ holds true, {\color{black}we have
\begin{align}
& TRCut(S, \bar{S}) = (\Theta_{G} - (\Theta_{S} + \Theta_{\bar{S}}))
\left(\frac{1}{|S|}+\frac{1}{|\bar{S}|}\right). \nonumber
\end{align}
For a given graph $G$}, the graph mutuality tendency $\Theta_{G}$ is
a constant, the minimization problem in eq.(\ref{eq:SP-P}) is
equivalent to the following maximization {\color{black}problem:
\begin{align}
&\max_{S} \left\{(\Theta_{S} +
\Theta_{\bar{S}}-\Theta_G)\left(\frac{1}{|S|}+\frac{1}{|\bar{S}|}\right)\right\}\label{eq:SP-P2}
\end{align}
Hence}, minimizing the cross-cluster mutuality tendency is
equivalent to maximize the within-cluster mutuality tendency. Using
the results presented in Theorem~\ref{thm:Laplacian}, we prove the
following theorem which provides the solution to the above mutuality
tendency optimization problem.

\begin{theorem}[]\label{thm:solve}
Given the tendency Laplacian matrix $L_T=D_T-T$, the signs of the
eigenvector of $L_T$ corresponding to the smallest non-zero
eigenvalue indicate the optimal solution $(S,\bar{S})$ to the
optimization problem eq.(\ref{eq:SP-P}).
\end{theorem}
\begin{Proofwithheader}{}
Define the column vector $f_S=[f_S(1),\cdots, f_S(n)]^T$ with
respect to a partition $S\cup \bar{S}=V$ as follows:
\begin{equation}
f_S(i) = \left\{
\begin{array}{ll}
    \sqrt{|\bar{S}|/|S|} &\textrm{if $i\in S$}\\
    -\sqrt{|{S}|/|{\bar{S}}|} &\textrm{if $i\in \bar{S}$}\\
\end{array}
\right..\label{eq:fsp}
\end{equation}
Then, by applying
Theorem~\ref{thm:Laplacian}, we have
\begin{align}
f_S^TL_Tf_S & = \sum_{i\sim j} {T_{ij}(f_S(i)-f_S(j))^2} \nonumber \\
& = (\frac{|S|}{|\bar{S}|}+\frac{|\bar{S}|}{|S|}+2)\sum_{i\in S\sim j\in \bar{S}} {T_{ij}} \nonumber \\
& = |V|(m^{(\omega)}_{\partial S} - {m}^{(\mu)}_{\partial
S})(\frac{1}{|\bar{S}|}+\frac{1}{|S|}),\nonumber \\
& = |V|\Theta_{\partial
S}(\frac{1}{|\bar{S}|}+\frac{1}{|S|}).\label{eq:fsptm1}
\end{align}
In addition, we have $f_S^Tf_S = \|f_S\|^2=|V|$. Hence,
Rayleigh-quoient for $L_T$ is
\begin{align}
\frac{f_S^TL_Tf_S}{f_S^Tf_S}&=TRCut(S,\bar{S}) \geq \lambda(L_T),\nonumber
\end{align}
where $\lambda(L_T)$ is the smallest non-zero eigenvalue of $L_T$.
Here $\lambda(L_T)$ cannot be $0$, because we have the constraint
$f_S\perp \textbf{1}$. From Theorem~\ref{thm:Laplacian},
$\textbf{1}$ is an eigenvector associated with eigenvalue $0$.
Hence, the problem of minimizing eq.(\ref{eq:SP-P}) can be
equivalently rewritten as
\begin{align}
&\min_{S} f_S^T{L}_Tf_S,
\mbox{ s.t.: $f_S \bot \textbf{1}$ in form of eq.~(\ref{eq:fsp}),
$\|f_S\|^2=|V|$.} \nonumber
\end{align}
Since the entries of the solution vector $f_S$ are only allowed to
take values in form of eq.(\ref{eq:fsp}), this is a discrete
optimization problem, which is known to be NP
hard~\cite{Tutorial-SC}. By relaxing the discreteness condition and
allowing $f_S(i)$ to take arbitrary values in $\mathbb{R}$, we have
the following relaxed optimization problem.
\begin{align}
&\min_{f_S\in \mathbb{R}^n} f_S^T{L}_Tf_S, 
&\mbox{ s.t.: $f_S \bot \textbf{1}$, and $\|f_S\|^2=|V|$.} \nonumber
\end{align}
The solution to this problem, i.e., the vector $f_S$, is the
eigenvector corresponding to the smallest non-zero eigenvalue
$\lambda({L}_T)$. Hence, we can approximate the minimizer of
$TRCut(S,\bar{S})$ using the eigenvector corresponding to
$\lambda({L}_T)$. To obtain a partition of the graph, we need to
convert the real-valued solution vector $f_S$ of the relaxed problem
to an indicator vector. One way to do this~\cite{Tutorial-SC} is to
use the signs of $f_S$ as indicator function, where node $v_i\in S$,
if $f_S(i)\geq 0$, and $v_i\in \bar{S}$, otherwise.
\end{Proofwithheader}

\subsection{Mutuality-tendency-aware spectral clustering: $K>2$}
For the case of finding $K > 2$ clusters $S_1\cup \cdots \cup
S_K=V$, we define the indicator vectors $h_k
=(h_{1k},\cdots,h_{nk})$,
\begin{align}
h_{ik}&=\left\{
\begin{array}{ll}
    \frac{1}{\sqrt{|S_k|}} & \textrm{if $v_i\in {S}_k$} \\
    0 & \textrm{otherwise}
\end{array}
\right.\label{eq:hij}
\end{align}
where $i=1,\cdots,n$ and $k=1,\cdots,K$. Let $H$ denote the
indicator matrix containing those $K$ indicator vectors as columns.
Observe that $H^TH = I$, $h_k^Th_k = 1$, and
\begin{align}
&h_k^T {L}_Th_k = \frac{\Theta_{\partial S_k}}{|S_k|}.
\label{eq:hLh}
\end{align}
Define the mutuality tendency ratio cut $TRCut(S_1, \cdots, S_K)$
for $K>2$ clusters as follows:
\begin{align}
&TRCut(S_1,\cdots,S_K)=\sum_{k=1}^K{\frac{\Theta_{\partial S_k}}{|S_k|}},\label{eq:hLh2}
\end{align}
where the ratio cut reduces to eq.(\ref{eq:TendCutF}) when $K=2$.
The problem of minimizing $TRCut$ can be formulated as
\begin{align}
&\min_{S_1,\cdots,S_K} TRCut(S_1, \cdots, S_K) =\min_{S_1,\cdots,S_K} tr(H^T{L}_TH) \nonumber \\
&\mbox{s.t.: $H^TH=I$, where $H$ is defined in eq.~(\ref{eq:hLh}).}
\nonumber
\end{align}
One way of solving this problem is utilizing the method used
in~\cite{Tutorial-SC} by relaxing the discreteness condition to have
a standard trace minimization problem as
\begin{align}
&\min_{H\in \mathbb{R}^{|V|\times K}} tr(H^T{L}_TH), &\mbox{ s.t.:
$H^TH=I$} \nonumber
\end{align}
The optimal solution $H$ contains the first $K$
eigenvectors of ${L}_T$ as columns. The clusters can be then
obtained by applying the K-means algorithm on those $K$
eigenvectors. The solution obtained minimizes the mutuality tendency
across clusters (which is equivalent to maximizing the within-cluster
mutuality tendency).

\noindent{\textbf{Choice of $K$.}} We choose $K$, i.e., the total
number of clusters, using the eigengap heuristic~\cite{Tutorial-SC}.
Theorem~\ref{thm:Laplacian} shows that $L_T$ has all real
eigenvalues. Denote the eigenvalues of $L_T$ in an increasing order,
i.e., $\lambda_1\leq\cdots\leq\lambda_n$, The index of the largest
eigengap, namely, $K := \argmax_{2\leq K\leq n}(g(K)),$ where $g(K)
= \lambda_{K} - \lambda_{K-1}$, $K = 2,\cdots,n$, indicates how many
clusters there are
in the network. 

\vspace{-0mm}
\section{Evaluations}\label{sec:eva}
\vspace{-0mm} In this section, we evaluate the performance of the
\emph{mutuality-tendency-aware} spectral clustering method by
comparing it with various symmetrization methods based digraph
spectral clustering algorithms. We only present the comparison
results for the adjacency matrix symmetrization method, with
objective matrix as $\bar{A}=(A+A^T)/{2}$. For other settings, we
obtained similar results and omit them here. We will 1) first test
the performances using synthetic datasets, and then 2) apply our
method to real online network datasets, e.g., Slashdot social
network, and discover stable clusters with respect to mutual and
one-way connections.

\subsection{Synthetic datasets}

\begin{figure*}[htb]
\begin{center}
    \subfigure[(i)All edges,
(ii)Bidirectional edges, (iii)Unidirectional
edges]{\hspace{-10pt}\includegraphics[width=0.5\textwidth]{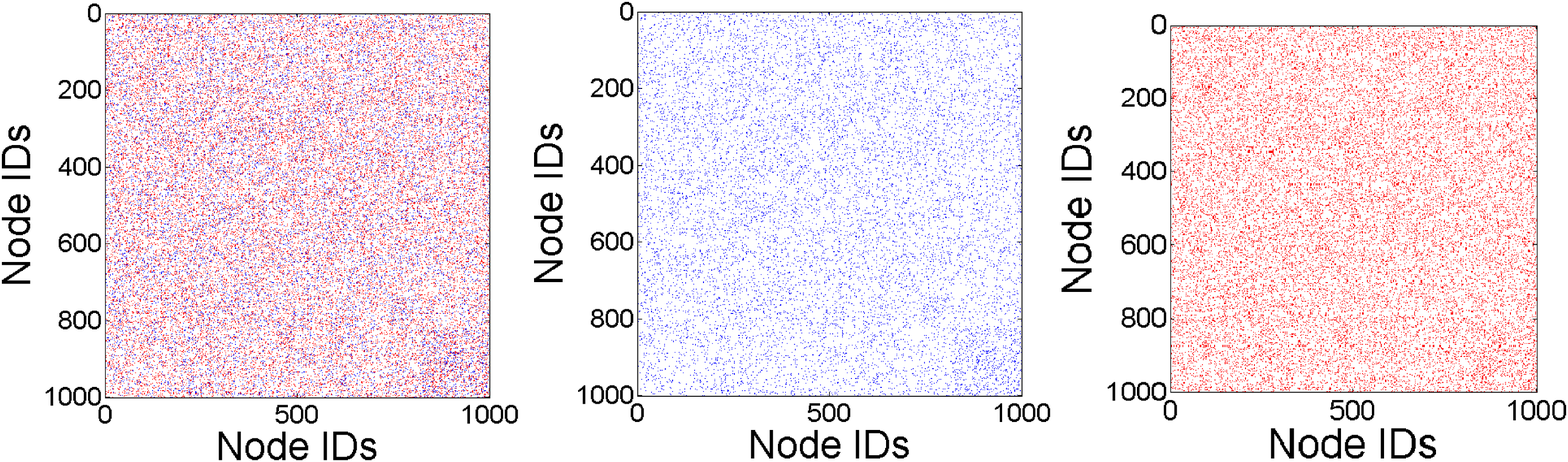}\label{fig:synfign01}}
    \subfigure[(i)All edges, (ii)Bidirectional edges, (iii)Unidirectional
edges]{\includegraphics[width=0.5\textwidth]{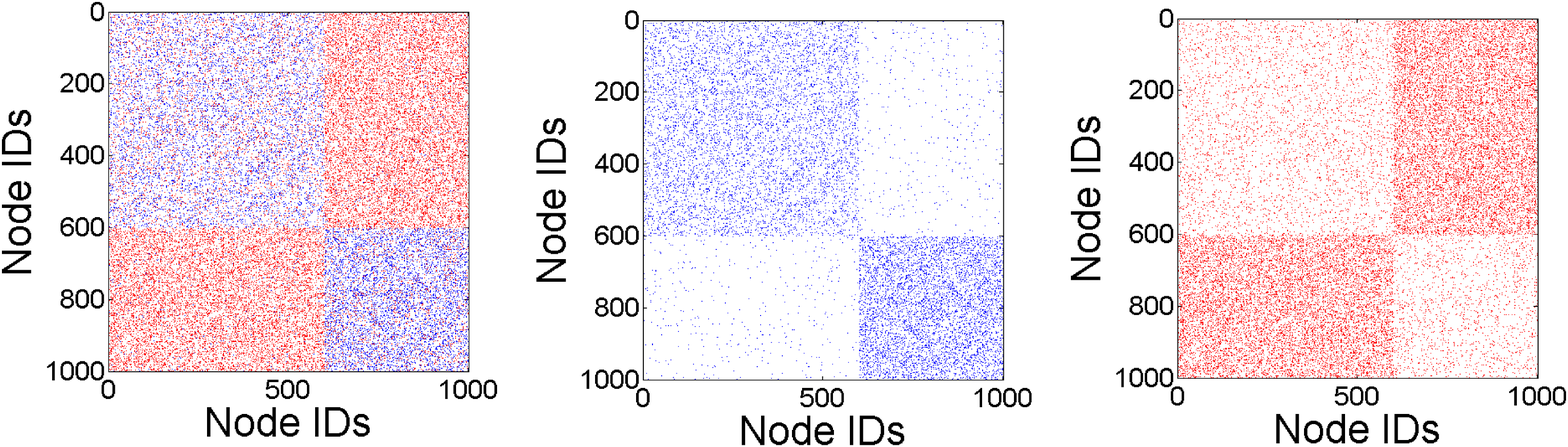}\label{fig:synfign33}}
\end{center}
\vspace{-0mm} \caption{Simulation results on synthetic dataset with
$K=2$ clusters. Fig.2(a)(i)-Fig.2(a)(iii) show the clusters detected
by traditional spectral clustering algorithm, and
Fig.2(b)(i)-Fig.2(b)(iii) show the clusters extracted using our
mutuality-tendency-aware spectral clustering
algorithm.}\label{fig:synthetic} \vspace{-0mm}
\end{figure*}


We first consider synthetic datasets designed specifically to test
the performance of our mutuality-tendency-aware spectral clustering
method. We randomly generate a network, with $1000$ nodes. There are
$38000$ directed edges (around $3.8\%$ of all directed node
pairs\footnote{As we observed from real social networks, e.g.,
Slashdot.com~\cite{Slashdotdata}, an online commenting network
dataset, which will be discussed in the next section, the sparsity
of its ``core'' network is around $0.19\%$. Here, we choose $3.8\%$,
that is $20$ times large of the real network sparsity, just for the
ease of visualization of the clustering structure.
}) in total, among which one third of them around $12666$ edges are
bidirectional, and two third of them around $25334$ edges are
unidirectional. Those nodes fall into $2$ clusters, with $600$ and
$400$ nodes respectively, where around $93.5\%$ of the bidirectional
edges are randomly placed \emph{within} clusters, and around
$80.8\%$ of the unidirectional edges are randomly placed
\emph{across} clusters.

We show in Fig.~\ref{fig:synfign01}(i)-Fig.~\ref{fig:synfign01}(iii)
that the traditional spectral clustering algorithm with
$\bar{A}=(A+A^T)/2$ as the objective results in clusters with $180$
and $820$ nodes respectively, which does not reflect the underlying
structure (See
Fig.~\ref{fig:synfign01}(i)-Fig.~\ref{fig:synfign01}(iii), because
it clusters nodes without considering the stability difference
between mutual connections and one-way connections.
On the other hand, our proposed mutuality-tendency-aware spectral
clustering method can cluster the nodes into groups with exactly
$600$ and $400$ nodes (See
Fig.~\ref{fig:synfign33}(i)-Fig.~\ref{fig:synfign33}(iii)), which
clearly group nodes with more mutual (stable) connections together
and separate nodes connected via one-way connections.

\vspace{-0mm}
\begin{table}[!htb]
\centering \caption{Ave. mutuality tendency comparison on synthetic
dataset}\label{tab::tendency}
\begin{tabular}{|p{3.0cm}||c|c|c|c|}
    \hline
. &  $\theta_G$    &   $\theta_S$  & $\theta_{\bar{S}}$   & $\theta_{\partial S}$  \\
    \hline
Tendency aware clustering   & 0.0112 & 0.0172 & 0.0314  & 8.25e-5 \\
Traditional clustering   & 0.0112 & 0.0115 & 0.0202  & 0.0096\\
    \hline
\end{tabular}
\vspace{-0mm}
\end{table}

\begin{figure*}[htb]
\begin{center}
    \subfigure[(i)All edges,
(ii)Bidirectional edges, (iii)Unidirectional
edges]{\hspace{-10pt}\includegraphics[width=0.5\textwidth]{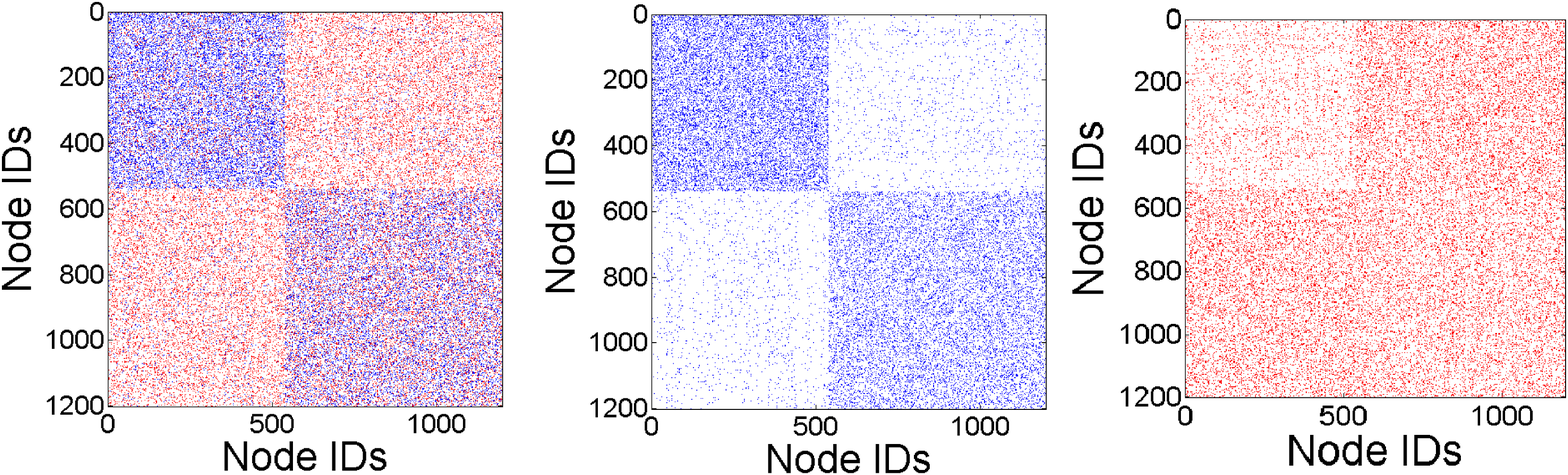}\label{fig:synfign011}}
    \subfigure[(i)All edges, (ii)Bidirectional edges, (iii)Unidirectional
edges]{\includegraphics[width=0.5\textwidth]{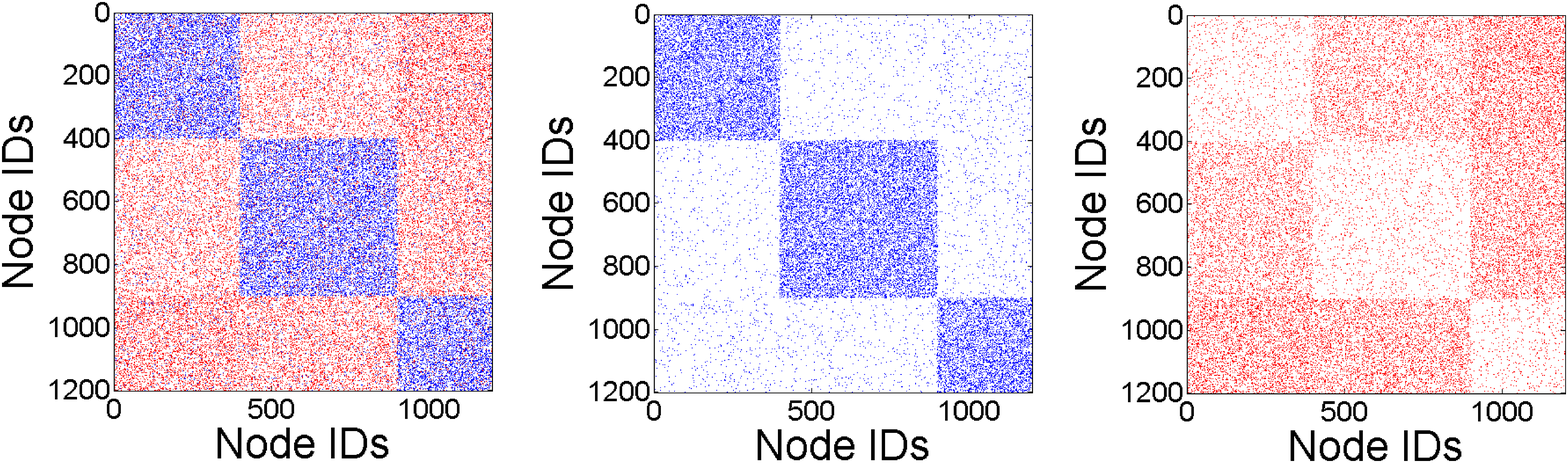}\label{fig:synfign331}}
\end{center}
\vspace{-0mm} \caption{This synthetic dataset is generated in $K=3$
clusters, with $500$, $400$ and $300$ nodes, respectively. There are
54675 directional edges, among which 27336 edges are bidirectional
and 27339 edges are unidirectional. We are randomly placed $90.02\%$
of the bidirectional edges in clusters, and $89.6\%$ of the
unidirectional edges across clusters.
Fig.3(a)(i)-Fig.3(a)(iii) show that traditional spectral clustering
algorithm detects clusters with $661$, $538$ and $1$ entities,
respectively, while our method identify correct clusters (See
Fig.3(b)(i)-Fig.3(b)(iii)).}\label{fig:synthetic2} \vspace{-0mm}
\end{figure*}

{\color{black}Furthermore, given the cluster mutuality tendency
$\Theta_{S}$, we denote the average mutuality tendency of $S$ as
$\theta_{S}=\Theta_{S}/N_S$, with $N_S=|S|(|S|-1)/2$ as the total
number of dyads in $S$. Similarly, we have the average mutuality
tendency of $G$, $\bar{S}$, and $\partial S$ as
$\theta_{G}=\Theta_{G}/N_d$,
$\theta_{\bar{S}}=\Theta_{\bar{S}}/N_{\bar{S}}$, and
$\theta_{\partial{S}}=\Theta_{\partial{S}}/(|S||\bar{S}|)$,
respectively.} Table~\ref{tab::tendency} shows the average mutuality
tendencies of the cluster results obtained by two methods, where we
can see that the \emph{mutuality-tendency-aware} spectral clustering
algorithm can group nodes together with higher within-cluster
tendencies than that by traditional spectral clustering. On the
other hand, the cross-cluster tendency obtained using our method is
very close to $0$, which means that the dyads across the clusters
establish the mutual connections without any tendency (or purely
independently). In addition,
we generated synthetic dataset with $K>2$ clusters, and similar results are obtained shown in Fig.~\ref{fig:synthetic2}. 



\vspace{-0mm}
\subsection{Real Social Networks}
\vspace{-0mm} In the second set of simulations, we applied our
\emph{mutuality-tendency-aware} spectral clustering algorithm to
several real social network datasets, e.g.,
Slashdot~\cite{Slashdotdata}, Epinions~\cite{richardson2003trust},
and email communication network~\cite{leskovec2007graph} datasets,
and compare with various symmetrization methods based digraph
clustering algorithms, such as $A=(A+A^T)/2$, $AA^T$ and
$F_{\pi}=\Pi P$. Here we only show the comparison results with
adjacency matrix symmetrization based digraph spectral clustering on
Slashdot dataset. All other settings lead to similar results and we
omit them here for brevity.

Slashdot is a technology-related news website founded in 1997. Users
can submit stories and it allows other users to comment on them. In
2002, Slashdot introduced the Slashdot Zoo feature which allows
users to tag each other as friends or foes. The network data we used
is the Slashdot social relation network, where a directed edge from
$i$ to $j$ indicates an interest from $i$ to $j$'s stories (or
topics). Hence, two people with mutual connections thus share some
common interests, while one-way connections infer that one is
interested in the other's posts, but the interests are not
reciprocated back. The Slashdot social network data was collected
and released by Leskovec~\cite{Slashdotdata} in November 2008.

%
%

\vspace*{-0.0cm}
\begin{table}[!htb]
\centering \caption{Statistics of Slashdot social network
Dataset}\label{Tab:Sla} {
\begin{tabular}{|c|c|}
\hline Nodes & $77360$\\
\hline Edges & $828161$\\
\hline Unidirectional edges & $110199$\\
\hline Bidirectional edges & $717962$\\
\hline Nodes in largest SCC & $ 70355$\\
\hline Edges in largest SCC & $ 818310$\\
\hline Unidirectional edges in largest SCC & $100930$\\
\hline Bidirectional edges in largest SCC & $717380$\\
\hline Nodes in the ``core'' component & $10131$\\
\hline Edges in the ``core'' component & $197378$\\
\hline Unidirectional edges in the ``core'' component & $21404$\\
\hline Bidirectional edges in the ``core'' component & $175974$\\
\hline\end{tabular}} \vspace*{-0.0cm}
\end{table}

\begin{figure*}[htb]
\begin{center}
    \subfigure[(i)All edges, (ii)Bidirectional edges, (iii)Unidirectional edges]{\hspace{-10pt}\includegraphics[width=0.5\textwidth]{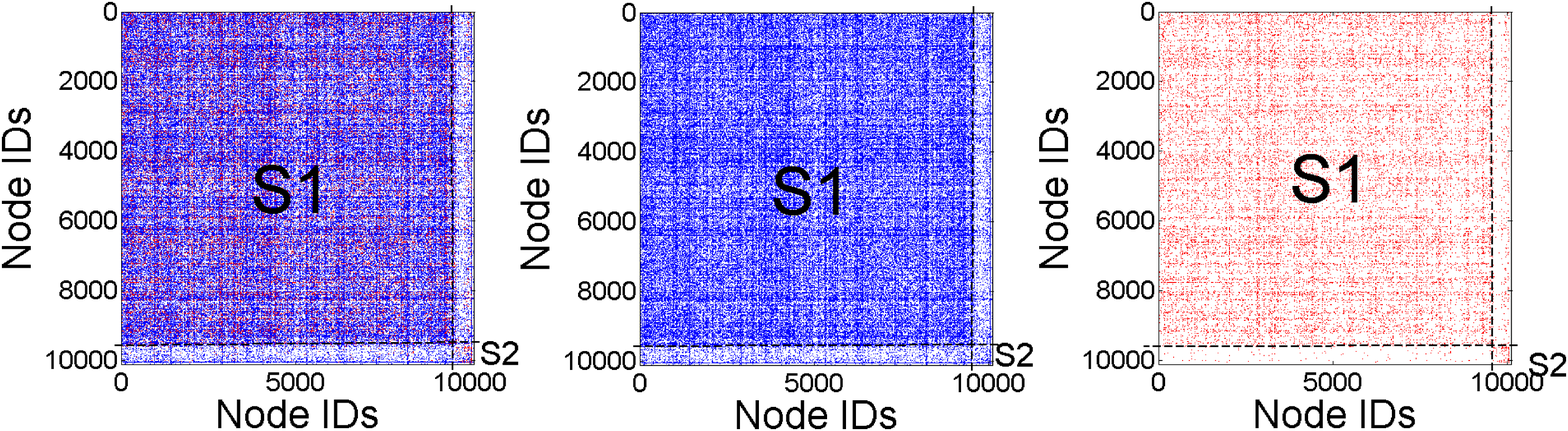}\label{fig:Sla14}}
    \subfigure[(i)All edges, (ii)Bidirectional edges, (iii)Unidirectional
edges]{\includegraphics[width=0.5\textwidth]{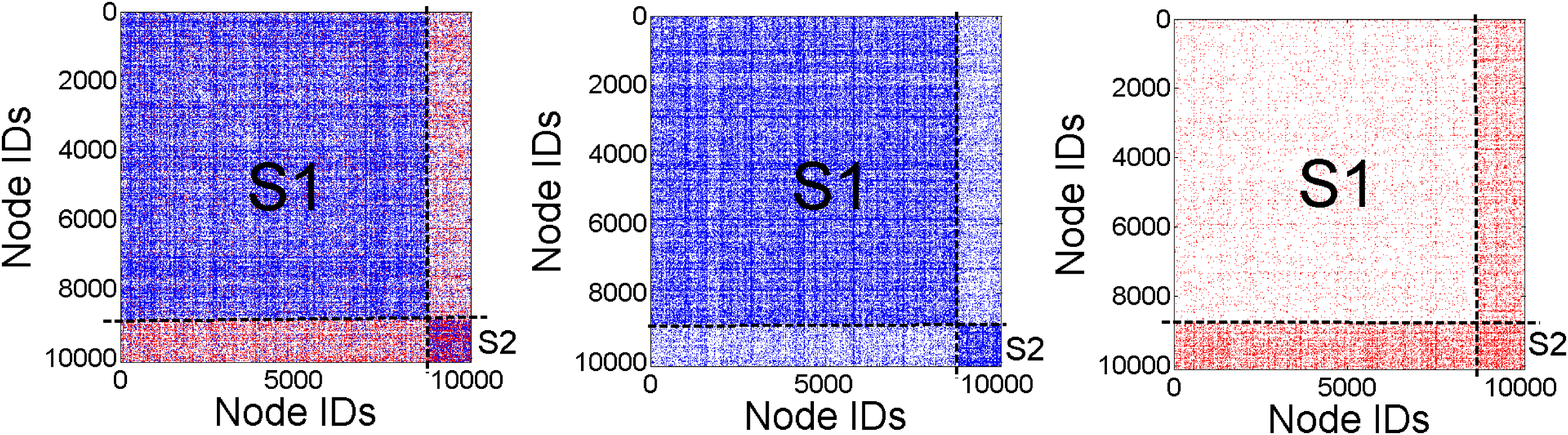}\label{fig:Sla00}}
\end{center}
\vspace{-0mm} \caption{Simulation results on Slashdot social network
dataset. Fig.4(a)(i)-Fig.4(a)(iii) show the clusters detected by
traditional spectral clustering algorithm, and
Fig.4(b)(i)-Fig.4(b)(iii) show the clusters extracted using our
mutuality-tendency-aware spectral clustering
algorithm.}\label{fig:Slashdot} \vspace{-0mm}
\end{figure*}

The statistics\footnote{Here, the total number of edges is smaller than that is shown on the website~\cite{Slashdotdata}, because we do not count for those selfloops.}
are shown in Table~\ref{Tab:Sla}. It shows that the largest strongly connected component (SCC)
include about $70355$ nodes. Then, we remove those nodes with very low in-degrees
and out-degrees, say no more than or equal to $2$. By finding the largest strongly connected component of the remaining graph, we extract a ``core'' of the network with $10131$ nodes and $197378$ edges, among which there are $21404$ unidirectional edges and $175974$ bidirectional edges, respectively.

In our evaluations, we observe that there is a large ``core'' of the network, 
and all other users are attached to this core network. In our study,
we are interested in extracting the community structure from the
``core'' network.

When applying our spectral clustering algorithm to the ``core''
network, two clusters with $8892$ and $1239$ nodes are detected
(shown in Fig.\ref{fig:Sla00}(i)-Fig.\ref{fig:Sla00}(iii)). In our
result, a large portion (about $35.04\%$) of cross-cluster edges are
unidirectional edges which in turn yield lower mutuality tendency
across clusters. On the other hand, when using the traditional
symmetrized $\bar{A}=(A+A^T)/{2}$, two clusters with $9640$ and
$491$ nodes are extracted instead (shown in
Fig.\ref{fig:Sla14}(i)-Fig.\ref{fig:Sla14}(iii)). We can see that
the clustering result obtained using the traditional spectral
clustering method has only around $5.75\%$ of the total edges across
clusters as unidirectional edges, which boost up
the mutuality tendency across clusters. 
However, in our clustering result, we have more unidirectional edges
placed across clusters, which decreases the mutuality tendency
across clusters. From Fig.~\ref{fig:Sla00}(i), we can clearly see
that we have unidirectional (red) edges dominating the cross-cluster
parts.

Table~\ref{tab::Sla_tendency} shows the average mutuality tendency comparison between different clustering methods, where 
we can see that the mutuality-tendency-aware spectral clustering
algorithm can group nodes together with higher within-cluster
tendencies than that of traditional spectral clustering.
\vspace{-0mm}
\begin{table}[!htb]
\centering \caption{Ave. mutuality tendency comparison on Slashdot
dataset}\label{tab::Sla_tendency}
\begin{tabular}{|p{3.0cm}||c|c|c|c|}
    \hline
. &  $\theta_G$ & $\theta_{S1}$  & $\theta_{S2}$   & $\theta_{\partial S}$  \\
    \hline
Tendency aware clustering   & 0.0017 & 0.0049 & 0.0028  & 0.00033 \\
Traditional clustering   & 0.0017 & 0.0018 & 0.0021  & 0.00070\\
    \hline
\end{tabular}
\vspace{-0mm}
\end{table}

\vspace{-0mm}\section{Conclusion}\label{sec:conclusion}
In this paper, we 
establish a generalized mutuality tendency theory to capture the
tendencies of clustered node pairs to establish mutual connections
more frequently than those occur by chance. Based on our mutuality
tendency theory, we develop a mutuality-tendency-aware spectral
clustering algorithm that can detect stable clusters, by maximizing
the within-cluster mutuality tendency and minimizing the
cross-cluster mutual tendency. Extensive simulation results on
synthetic, and real online social network datasets, such as
Slashdot, demonstrate that our proposed
\emph{mutuality-tendency-aware} spectral clustering method resolves
more stable social community structures than traditional spectral
clustering methods.\vspace{-0mm}


\bibliographystyle{abbrv}
\bibliography{Yanhua-long-bak}

\begin{thebibliography}{10}

\bibitem{berscheid2005psychology}
E.~Berscheid and P.~Regan.
\newblock {\em The psychology of interpersonal relationships}.
\newblock Pearson Prentice Hall, 2005.

\bibitem{descioli2011best}
P.~DeScioli, R.~Kurzban, E.~Koch, and D.~Liben-Nowell.
\newblock Best friends: Alliances, friend ranking, and the myspace social
  network.
\newblock {\em Perspectives on Psychological Science}, 6(1):6--8, 2011.

\bibitem{golder2009structural}
S.~Golder, S.~Yardi, and A.~Marwick.
\newblock A structural approach to contact recommendations in online social
  networks.
\newblock {\em Workshop on Search in Social Media (SSM)}, 2009.

\bibitem{gouldner1960norm}
A.~Gouldner.
\newblock The norm of reciprocity: A preliminary statement.
\newblock {\em American sociological review}, pages 161--178, 1960.

\bibitem{hagen1992new}
L.~Hagen and A.~Kahng.
\newblock New spectral methods for ratio cut partitioning and clustering.
\newblock {\em IEEE Transactions on Computer-Aided Design of Integrated
  Circuits and Systems}, 11(9):1074--1085, 1992.

\bibitem{heider1946attitudes}
F.~Heider.
\newblock Attitudes and cognitive organization.
\newblock {\em Journal of psychology}, 21(1):107--112, 1946.

\bibitem{jamali2011modeling}
M.~Jamali, G.~Haffari, and M.~Ester.
\newblock Modeling the temporal dynamics of social rating networks using
  bidirectional effects of social relations and rating patterns.
\newblock In {\em WWW}, 2011.

\bibitem{katz1955measurement}
L.~Katz and J.~Powell.
\newblock Measurement of the tendency toward reciprocation of choice.
\newblock {\em Sociometry}, 18(4):403--409, 1955.

\bibitem{kurucz2007spectral}
M.~Kurucz, A.~Benczur, K.~Csalogany, and L.~Lukacs.
\newblock Spectral clustering in telephone call graphs.
\newblock In {\em WebKDD 2007}.

\bibitem{kwak2011fragile}
H.~Kwak, H.~Chun, and S.~Moon.
\newblock Fragile online relationship: A first look at unfollow dynamics in
  twitter.
\newblock In {\em CHI}, 2011.

\bibitem{leicht2008community}
E.~Leicht and M.~Newman.
\newblock Community structure in directed networks.
\newblock {\em Physical Review Letters}, 100(11):118703, 2008.

\bibitem{leskovec2007graph}
J.~Leskovec, J.~Kleinberg, and C.~Faloutsos.
\newblock Graph evolution: Densification and shrinking diameters.
\newblock {\em ACM Transactions on Knowledge Discovery from Data (TKDD)},
  1(1):2--es, 2007.

\bibitem{leskovec2010empirical}
J.~Leskovec, K.~Lang, and M.~Mahoney.
\newblock Empirical comparison of algorithms for network community detection.
\newblock In {\em WWW}, 2010.

\bibitem{WAW2010Yanhua}
Y.~Li and Z.-L. Zhang.
\newblock Random walks on digraphs, the generalized digraph laplacian and the
  degree of asymmetry.
\newblock In {\em LNCS WAW 2010}, Stanford, CA, 2010. LNCS.

\bibitem{Luxburg:2007}
U.~Luxburg.
\newblock A tutorial on spectral clustering.
\newblock {\em Statistics and Computing}, 17:395--416, December 2007.

\bibitem{miller1972structural}
H.~Miller and D.~Geller.
\newblock Structural balance in dyads.
\newblock {\em Journal of Personality and Social Psychology}, 21(2):135, 1972.

\bibitem{mishra2007clustering}
N.~Mishra, R.~Schreiber, I.~Stanton, and R.~Tarjan.
\newblock Clustering social networks.
\newblock {\em Algorithms and Models for the Web-Graph, 2007}.

\bibitem{price1966psychological}
K.~Price, E.~Harburg, and T.~Newcomb.
\newblock Psychological balance in situations of negative interpersonal
  attitudes.
\newblock {\em Journal of Personality and Social Psychology}, 3(3):265, 1966.

\bibitem{richardson2003trust}
M.~Richardson, R.~Agrawal, and P.~Domingos.
\newblock Trust management for the semantic web.
\newblock {\em The SemanticWeb-ISWC}, 2003.

\bibitem{rodrigues1967effects}
A.~Roudrigues.
\newblock Effects of balance, positivity, and agreement in triadic social
  relations.
\newblock {\em Journal of Personality and Social Psychology}, 5(4):472, 1967.

\bibitem{edbt:SatuluriP11}
V.~Satuluri and S.~Parthasarathy.
\newblock Symmetrizations for clustering directed graphs.
\newblock In {\em EDBT/ICDT}, 2011.

\bibitem{pami:ShiM00}
J.~Shi and J.~Malik.
\newblock Normalized cuts and image segmentation.
\newblock {\em IEEE Trans. Pattern Anal. Mach. Intell.}, 22(8):888--905, 2000.

\bibitem{Slashdotdata}
Slashdot.
\newblock dataset.
\newblock {\em http://snap.stanford.edu/data/soc-Slashdot0811.html}.

\bibitem{smyth2005spectral}
S.~Smyth.
\newblock A spectral clustering approach to finding communities in graphs.
\newblock In {\em SDM 2005}.

\bibitem{Tutorial-SC}
U.~von Luxburg.
\newblock A tutorial on spectral clustering.
\newblock Technical Report No.TR-149, Max Planck Institute for Biological
  Cybernetics, 2006.

\bibitem{WangD:KDD10}
X.~Wang and I.~Davidson.
\newblock Flexible constrained spectral clustering.
\newblock In {\em KDD 2010}, pages 563--572, 2010.

\bibitem{WestonLIZEN05}
J.~Weston, C.~S. Leslie, E.~Ie, D.~Zhou, A.~Elisseeff, and W.~S. Noble.
\newblock Semi-supervised protein classification using cluster kernels.
\newblock {\em Bioinformatics}, 21(15):3241--3247, 2005.

\bibitem{wolfe1997social}
A.~Wolfe.
\newblock Social network analysis: Methods and applications.
\newblock {\em American Ethnologist}, 24(1):219--220, 1997.

\bibitem{icmlZhouHS05}
D.~Zhou, J.~Huang, and B.~Sch{\"o}lkopf.
\newblock Learning from labeled and unlabeled data on a directed graph.
\newblock In {\em ICML}, 2005.

\end{thebibliography}
\end{document}